\documentclass[11pt]{article}
\pdfoutput=1
\usepackage[margin=1in]{geometry}
\usepackage{mathptmx}
\usepackage{titlesec}
\usepackage{etoolbox}
\usepackage{setspace}
\usepackage{graphicx}
\usepackage{natbib}
\usepackage{newtxtext}
\usepackage{newtxmath}
\usepackage{hyperref}
\usepackage{float}
\usepackage{soul}
\hypersetup{
	colorlinks = true,
	urlcolor   = blue,
	citecolor  = black,
}

\newtheorem{proposition}{Proposition}
\newtheorem{proof}{Proof}
\newcommand{\RomanNumeralCaps}[1]
\linenumbers

\usepackage{xcolor}

\title{Maximum Likelihood Filtering for Particle Tracking in Turbulent Flows}

\begin{document}
	\maketitle
	Griffin M. Kearney$\mathrm{^{a,b}}$\footnote{griffin.kearney@opbdatainsights.com}, Kasey M. Laurent$\mathrm{^{b}}$\footnote{klaurent@syr.edu}, Reece V. Kearney
	
	$\mathrm{^a}${\it OpB Data Insights LLC, Syracuse, NY 13224}
	
	$\mathrm{^b}${\it Syracuse University, Syracuse, NY 13244}

	\vskip\bigskipamount 
	\vskip\medskipamount
	\leaders\vrule width \textwidth\vskip0.4pt 
	\vskip\bigskipamount 
	\vskip\medskipamount
	\nointerlineskip
	\begin{abstract}
		Lagrangian Particle Tracking (LPT) enables practitioners to study various concepts in turbulence by measuring particle positions in flows of interest.
		This data is subject to measurement errors, and filtering techniques are applied to mitigate these errors and improve the accuracy of analyses utilizing the data.
		We develop a new type of position filter through use of maximum likelihood estimation by considering both measurement errors and stochastic process physics.
		The maximum likelihood estimation scheme we develop is general and can accommodate many different stochastic process models, enabling it to be applied to many different turbulent flows. 
		In this work, we propose a process model similar to existing, complimentary work in the development of B-splines.
		We compare our filtering scheme to existing schemes and find that our filter out performs the scheme proposed by \cite{Mordant2004} considerably, and produces similar performance to spline filters, proposed by \cite{Gesemann2018}.
		In comparing to the latter, we note that the maximum likelihood treatment provides a general framework which is capable of producing different filters based on the physics of interest, whereas the spline filters are built on less specific filtering theory and are therefore more difficult to adapt across diverse use cases in fluids.
		We quantify the performance of each of the filtering methods using error metrics which consider both noise reduction as well as signal degradation, and together these are used to define a concept of filter efficiency.
		The maximum likelihood filter developed in this work is shown to be the most efficient among all the methods examined when applied to simulated isotropic turbulence data from the Johns Hopkins Database.
	\end{abstract}
	
	\section{Introduction}
	\label{sec:intro}
	The formation of rain in clouds, the process of turbulent combustion, and the dispersion of pollutants in the atmosphere are all complex, 3D phenomena often evaluated in the Lagrangian frame \cite{Saffman1956,Celis2015,Weil1992}. 
	In experiments, Lagrangian particle tracking (LPT) techniques, which involve optically tracking individual particles within the flow of interest, enable the evaluation of Lagrangian statistics. 
	Particle positions are used to reconstruct trajectories which, in turn, can be differentiated to estimate the velocity and acceleration of the particles. 
	By considering these higher-order derivatives, specifically within turbulent flows, one can gain a fundamental understanding of the underlying physics that drive various flow phenomena.
	
	Estimating particle accelerations in turbulence is challenging and requires highly accurate measurements of the particle positions. 
	Uncertainty in these positions has a detrimental impact on estimates for the acceleration. 
	Differentiation increases the uncertainty in the signal by a factor inversely proportional to the time step, compounding through higher derivatives.
	This uncertainty can be minimized by increasing the spatial resolution, and by applying the appropriate filters to remove noise from the signal. 
	The spatial resolution is limited based on both the imaging system and the particle size, but can be increased to sub-pixel accuracy \cite{Kaehler2012}. 
	To mitigate the effect of noise from the measurements, in addition to optimizing the imaging system, it is essential to choose an appropriate filter and filter scale.
	Failing to do so can have a significant impact on the estimated acceleration statistics \cite{Lawson2018}. 
	In real-world data collection, measurement noise is unavoidable, and finding methods to reduce the impact of these errors is beneficial to data analysis.
	
	A common method for computing the Lagrangian velocity and acceleration statistics from noisy position data involves the convolution of data with a Gaussian filter kernel. 
	The details of this technique are described in \cite{Mordant2004}. 
	A major limitation of this approach is its tendency towards selection bias - velocity and acceleration estimates are undefined when there are gaps in the data, and the kernel cannot be properly applied to the ends of tracks, causing shorter tracks to be discarded. 
	These shorter tracks are often associated with faster particles, which remain in the interrogation window for relatively short periods of time. 
	Since these particles are more likely to be associated with larger accelerations \cite{Sawford2003, Voth2002}, ignoring them can result in the tails of the acceleration probability density function (pdf) to be suppressed \cite{Crawford2004, Ouellette2006}.
	
	One way to overcome this drawback is to significantly oversample the position of the particles. 
	By oversampling the position, the random position errors can be minimized with a large filter support, which in turn provides a more accurate estimate for the temporal derivatives of the particle's position. 
	Fluid particles in turbulent flows exhibit extreme accelerations more frequently than what is predicted by a Gaussian distribution. 
	These large accelerations, combined with the need to oversample to minimize bias, leads to sample periods sometimes exceeding 100 Kolmogorov timescales, corresponding to tens of thousands of samples per second \cite{Voth2002}.
	It is difficult and costly to achieve sampling rates this high while maintaining a large enough volume of observation to capture phenomena of interest.
	
	Another approach to finding the Lagrangian velocity and acceleration from noisy position data was recently proposed by \cite{Gesemann2018}, which fits position data with penalized cubic B-splines. 
	This method avoids the selection biases of the Gaussian filter method by enabling interpolation for missing data and minimizing the end effects. 
	Additionally, the estimates for the acceleration variance and flatness are less dependent on the filter length compared to the Gaussian filter approach \cite{Lawson2018}.
	The work presented in \cite{Gesemann2018} relies heavily on filtering theory and the approximation of the commonly used Wiener filter. 
	In developing the procedure for computing the spline fits, the authors observe that, through appropriate choice of cubic spline weights, the technique closely matched the implementation of a Wiener filter.
	
	In this paper, we utilize Maximum Likelihood Estimation (MLE) techniques to develop a high-performance, physically grounded data smoothing scheme. 
	Maximum Likelihood Estimation (MLE) techniques have proven to be effective in related tracking applications in bio-physics \cite{Bullerjahn2021}. We develop a framework to apply these types of techniques to problems related to LPT in turbulent environments. 
	Applying MLE techniques requires a probability model of underlying system dynamics to produce robust solutions.
	By focusing our attention on the motion of particles in turbulent flows, we are able to utilize physical intuitions and existing knowledge to formulate physically motivated maximum likelihood schemes.
	
	This paper is organized as follows. We describe the technical development of a noise filtering scheme by analyzing stochastic physical models in Section \ref{sec:tech-dev}. 
	In Section \ref{sec:methods}, we describe the numerical simulations and methods for acquiring the particle tracks used to validate the MLE filtering scheme. 
	For the analysis in Section \ref{sec:results}, we compare the optimal MLE filter to the well-established techniques of local Gaussian kernel smoothing \cite{Mordant2004} and B-spline fittings \cite{Gesemann2018}, and explicitly demonstrate instances in which the MLE smoothing solution outperforms the Gaussian smoothing scheme.
	We show that the Maximum Likelihood Estimation filter developed in this work demonstrates performance that is nearly identical to the performance of the B-spline fitting techniques developed in \cite{Gesemann2018}.
	

	\section{Technical Development}
	\label{sec:tech-dev}
	We begin in Section \ref{subsec:prob-model} by proposing a probability-based model of the systems of interest.
	We consider both process-based randomness, induced by the stochastic nature of turbulent systems, and measurement errors, induced by noise that is unavoidable in real-world data collection systems.
	Utilizing the probability models allows us to explicitly formulate a governing MLE scheme in Section \ref{subsec:dev-mle}, and we analyze the solution of this governing problem in Section \ref{subsec:analysis-mle}.
	
	\subsection{Probabilistic Modeling of the System}
	\label{subsec:prob-model}
	Let the true trajectory of a fluid particle be represented by the unknown vector $x$, which is comprised of discrete time samples $t = 1, 2, ...T$.
	Our data smoothing scheme seeks to estimate $x$ given a set of noisy measurements contained in a discrete temporal vector $y$, collected through experiments.
	We formulate the governing model of the estimation scheme through application of Bayes Rule
	\begin{equation}
		\label{eqn:Bayes-MLE}
		\mathrm{pr}(y,x) = \mathrm{pr}(y|x)\mathrm{pr}(x).
	\end{equation}
	Bayes Rule states that the joint probability of $y$ and $x$ is equal to the product of the conditional probability of $y$ given $x$ and the probability of $x$.
	Furthermore, Bayes Rule applies to marginal probabilities, and computation of the joint probability density function is performed using this technique.
	In the following sections, we construct the joint pdf using equation \ref{eqn:Bayes-MLE} by considering both process noise, governing $\mathrm{pr}(x)$, and measurement error, governing $\mathrm{pr}(y|x)$.
	We begin by modeling $\mathrm{pr}(x)$ by treating $x$ as a position undergoing a stochastic process, since the motion of a particle in a turbulent flow will be chaotic in general.
	
	\subsubsection{Stochastic Forcing}
	\label{subsubsec:stoch-force}
	Suppose that a fluid particle of interest has mass $m$.
	We model the particle as a point mass and write Newton's second law as the governing equation of its motion
	\begin{equation}
		\label{eqn:newton-law}
		m a(t) = F(t),
	\end{equation}
	where $a(t)$ and $F(t)$ are the acceleration of the particle and the force exerted on the particle by the flow at time $t$, respectively.
	We denote the vector of accelerations of the particle in discrete time as $a$, where $a(t)$ denotes the scalar acceleration experienced by the fluid particle at discrete time $t$.
	In general, $a$ may not have the same dimension as $x$ since the computation of $a$ often requires computing component differences of $x$, and can therefore fail to provide estimates for all $t = 1, 2, ... T$.
	We assume that $F(t)$ is discrete, stochastic in nature,
	and that the rate at which we measure the particle positions is fast enough such that the accelerations are locally correlated in discrete time. 
	Consequently, the change in the forces experienced by the particle can be written as
	\begin{equation}
		\label{eqn:stoch-force}
		F(t) = F(t-1) + u(t),
	\end{equation}
	for each discrete time $t$, with $u(t)$ taken to be a random variable sampled from a Gaussian distribution with mean $0$ and variance $\sigma_F^2$, which we denote as $N(0,\sigma_F^2)$. 
	We note that this choice for $u(t)$ parallels the interpretation of the B-splines technique in \cite{Gesemann2018} as a Kalman-like filter with changes in the acceleration modeled as white noise. 
	In practice, the changes in the acceleration can be non-Gaussian, exhibiting elevated tails corresponding to greater likelihood of high acceleration events; statistical observations of the tails can depend on the sample rate of the data collection, as seen in the statistics of Lagrangian velocity differences \cite{Mordant2001}. 
	In our present treatment, the Gaussian distribution governing $u(t)$ may be substituted for distributions which more accurately model application specific forcing, without fundamentally changing the analysis.
	We rewrite equation \ref{eqn:stoch-force} as
	\begin{equation*}
		F(t) - F(t-1) = u(t)
	\end{equation*}
	and incorporate equation \ref{eqn:newton-law}, which yields
	\begin{equation*}
		a(t) - a(t-1) = \frac{1}{m}u(t).
	\end{equation*}
	Thus we note that under these assumptions, the finite difference of the acceleration at time $t$,
	\begin{equation}
		\label{eqn:diff-accel}
		\Delta a(t) = a(t) - a(t-1)
	\end{equation}
	is itself a normal random variable sampled from the distribution
	\begin{equation*}
		\label{eqn:del-acc-dist}
		\Delta a(t) \sim N\bigg(0,\frac{\sigma_F^2}{m^2}\bigg).
	\end{equation*}
	Assuming that $a$ contains $T_a$ time samples, this implies that $\Delta a$ contains $T_a-1$ time samples, since the difference in equation \ref{eqn:diff-accel} cannot be computed at the first data point.
	With this in mind, we are able to write the probability density function of the vector $\Delta a$ as
	\begin{equation}
		\label{eqn:stoch-process-pdf}
		\rho(\Delta a) = \bigg(\frac{m}{\sqrt{2 \pi} \sigma_F}\bigg)^{T_a-1} e^{-\frac{m^2}{2 \sigma_F^2}||\Delta a||^2}.
	\end{equation}
	The probability density function contained in equation \ref{eqn:stoch-process-pdf} is used to quantify process randomness in the MLE scheme of section \ref{subsec:dev-mle}.
	It is worth noting that this model of stochastic forcing naturally arrives at a distribution (equation \ref{eqn:stoch-process-pdf}) which penalizes large third order time derivatives of the particle position.
	While our framework is sufficiently general to accommodate many other types of distributions in lieu of this forcing model, the result of considering this specifically is consistent with the penalty function for fitting B-splines proposed in \cite{Gesemann2018}.
	We emphasize that our treatment is both more general and is developed on physical first principles as opposed to using empirical techniques in filtering theory. 
	
	We refine equation \ref{eqn:stoch-process-pdf} in Section \ref{subsec:dev-mle} using finite difference techniques to formulate the term $\mathrm{pr}(x)$ in Bayes Rule in terms of $x$ instead of $a$.
	However, we pause before developing the refinement to describe a common model for measurement error.
	Measurement error is utilized for the Bayesian term $\mathrm{pr}(y|x)$, and allows us to elegantly state the general estimation scheme in \ref{subsec:dev-mle} when we return to refining equation \ref{eqn:stoch-process-pdf}.
	
	\subsubsection{Measurement Error}
	\label{subsubsec:meas-error}
	We assume that the vector of discrete time measurements $y$ is of the form
	\begin{equation}
		\label{eqn:meas-sum}
		y(t) = x(t) + z(t)
	\end{equation}
	where $y(t)$ is the measured position at time $t$, $x(t)$ is the true position at time $t$, and $z(t)$ is the measurement noise on the $t$-th time sample.
	The measurement error is often due to the presence of many independent minor sources of error all acting together.
	Since $z(t)$ is thus treated as the superposition of these many errors, the central limit theorem suggests that it is reasonable to assume that $z(t)$ is normally distributed.
	In many cases, it is possible to perform a calibration process so that the mean of this normally distributed error is zero.
	Therefore, by rewriting equation \ref{eqn:meas-sum}, we observe
	\begin{equation*}
		\label{eqn:meas-noise}
		y(t) - x(t) = z(t) \sim N(0,\sigma^2),
	\end{equation*}
	where $\sigma^2$ is the variance of the measurement error.
	Assuming that there are $T$ measurements taken, we write the probability density function of the error vector $z = y - x$ as
	\begin{equation}
		\label{eqn:meas-pdf}
		\eta(y-x) = \bigg(\frac{1}{\sqrt{2 \pi} \sigma}\bigg)^T e^{-\frac{1}{2 \sigma^2}||y - x||^2}.
	\end{equation}
	Equation \ref{eqn:meas-pdf} is treated as the conditional probability of the measurement $y$, given that the noiseless trajectory is $x$.
	
	\subsection{Development of the Maximum Likelihood Estimator}
	\label{subsec:dev-mle}
	We are now nearly positioned to utilize Bayes Rule
	\begin{equation*}
		\mathrm{pr}(y,x) = \mathrm{pr}(y|x)\mathrm{pr}(x).
	\end{equation*}
	We immediately may substitute equation \ref{eqn:meas-pdf} for the conditional probability $\mathrm{pr}(y|x)$ in the Bayesian product, writing
	\begin{equation}
		\label{eqn:prob-y-given-x}
		\mathrm{pr}(y|x) = \eta(y-x) = \bigg(\frac{1}{\sqrt{2 \pi} \sigma}\bigg)^T e^{-\frac{1}{2 \sigma^2}||y - x||^2},
	\end{equation}
	as it is explicitly in terms of $y$ and $x$.
	
	Finding a workable form for $\mathrm{pr}(x)$ requires further refinement of equation \ref{eqn:stoch-process-pdf} to phrase it in terms of $x$, rather than $a$.
	This is done by applying a simple finite difference scheme to approximate the time derivatives.
	Suppose that $x$ is a $T$ component discrete time vector.
	The second derivative at times $t = 2,3,...,T-1$ is approximated with simple central differences, computed using
	\begin{equation}
		\label{eqn:snd-ordr-diff}
		a(t) = \frac{x(t+1) - 2 x(t) + x(t-1)}{\Delta t^2}	
	\end{equation}
	where $\Delta t$ is the elapsed time between consecutive time samples. 
	Here, $a$ is a $T_a$ component discrete time vector where $T_a = T - 2$.
	
	There are many different weighting schemes that have been developed for approximating second derivatives in addition to this one, including methods which allow for estimation of the derivatives at the instants $t=1$ and $t=T$.
	The method proposed by \cite{Mordant2004}, which we use for comparison of results in later sections, is an example of an alternative weighting scheme. 
	This method also produces acceleration estimates with $T_a < T$, but the value of $T_a$ is dependent on the choice of filter length.
	Regardless of the specific finite difference scheme, the operation of approximating the second derivative using the sampled points is generally a linear operation.
	This is critical, as it allows us to write
	\begin{equation*}
		a = A x,
	\end{equation*}
	where $A$ is a matrix defined by the specific differencing scheme.
	In the case of the finite difference scheme shown in equation \ref{eqn:snd-ordr-diff}, $A$ would contain $-2$ along its main diagonal with $1$ on its first upper and lower diagonals, all other components would be $0$, and its entirety would be scaled by $\frac{1}{\Delta t^2}$. 
	
	The computation to calculate $\Delta a$ using $a$ by equation \ref{eqn:diff-accel} is also a linear operation. The composition of linear operations is itself linear, we write
	\begin{equation*}
		\label{eqn:stoch-process-matrix}
		\Delta a = (\Delta A) x
	\end{equation*}
	using the matrix product $\Delta A$ where $\Delta$ is a $T-1 \times T$ matrix defined by
	\begin{equation}
		\label{eqn:delta-mat-def}
		\Delta_{kk} = -1, \hspace{0.25cm} \Delta_{k,k+1} = 1, \hspace{0.25cm} k=1,2,...,T-1,
	\end{equation}
	with all other components equal to $0$.
	
	Utilizing the above, we rewrite equation \ref{eqn:stoch-process-pdf}, using equation \ref{eqn:delta-mat-def} to substitute expressions, and derive
	\begin{equation}
		\label{eqn:prob-x}
		\mathrm{pr}(x) = \rho(\Delta a) = \rho(\Delta A x) = \bigg(\frac{m}{\sqrt{2 \pi} \sigma_F}\bigg)^{T_a-1} e^{-\frac{m^2}{2 \sigma_F^2}||\Delta A x||^2}.
	\end{equation}
	With equations \ref{eqn:prob-y-given-x} and \ref{eqn:prob-x}, we write the joint probability of $y$ and $x$ using equation \ref{eqn:Bayes-MLE} as
	\begin{equation*}
		\mathrm{pr}(y,x) = \bigg(\frac{m}{\sqrt{2 \pi} \sigma_F}\bigg)^{T_a-1} \bigg(\frac{1}{\sqrt{2 \pi} \sigma}\bigg)^T e^{-\big(\frac{1}{2 \sigma^2}||y - x||^2 +\frac{m^2}{2 \sigma_F^2}||\Delta A x||^2\big)}.
	\end{equation*}
	Estimation of $x$ is then performed by computing its value through maximizing the joint probability.
	This is the governing MLE problem and is stated as
	\begin{equation*}
		\label{eqn:mle-problem-raw}
		\underset{x}{\max} \hspace{0.25cm} \bigg(\frac{m}{\sqrt{2 \pi} \sigma_F}\bigg)^{T_a-1} \bigg(\frac{1}{\sqrt{2 \pi} \sigma}\bigg)^T e^{-\big(\frac{1}{2 \sigma^2}||y - x||^2 +\frac{m^2}{2 \sigma_F^2}||\Delta A x||^2\big)}.
	\end{equation*}
	Applying common simplifications, this problem is equivalent to
	\begin{equation}
		\label{eqn:mle-problem-simplified}
		\underset{x}{\min} \hspace{0.25cm} \frac{1}{2 \sigma^2}||y - x||^2 +\frac{m^2}{2 \sigma_F^2}||\Delta A x||^2.
	\end{equation}
	
	Equation \ref{eqn:mle-problem-simplified} is convex in general, and therefore permits a unique minimum.
	Moreover, the minimum is computed in closed form, and the solution is captured in the following Proposition.
	\begin{proposition}
		\label{prop:mle-solution}
		The optimization problem \ref{eqn:mle-problem-simplified} is solved for
		\begin{equation*}
			x = (I + \mu^2 A^T \Delta^T \Delta A)^{-1} y
		\end{equation*}
		with $\mu = \frac{m \sigma}{\sigma_F}$, $I$ denoting the identity matrix of appropriate size, and $A^T$, $\Delta^T$ denoting the transpose of $A$ and $\Delta$, respectively.
	\end{proposition}
	\noindent The proof of Proposition \ref{prop:mle-solution} is included in Appendix \ref{subsec:appen-proof}.
	
	As we move to deepen the understanding of the MLE solution, we interpret the matrix
	\begin{equation*}
		G(\mu) = (I + \mu^2 A^T \Delta^T \Delta A)^{-1}
	\end{equation*}
	as a discrete, one parameter ($\mu$) filter acting on the noisy measurement data.
	Pleasingly, the filter parameter has a natural physical interpretation, depending on the interplay of the particle's mass, the measurement noise levels, and the volatility of the stochastic process describing the dynamics.
	When considering tracer particles (or fluid particles), it is helpful to rewrite $\mu$ as
	\begin{equation*}
		\label{eqn:mu-std-accel}
		\mu = \frac{\sigma}{\frac{\sigma_F}{m}} = \frac{\sigma}{\sigma_a},
	\end{equation*} 
	where $\sigma_a$ is the variance of the change in the acceleration.
	The variance of the change in forces experienced by the particle, $\sigma_F$, is rewritten as $\sigma_F = m \sigma_a$, and the mass term is dropped from our definition of $\mu$. 
	Conversely, in this treatment we can look at inertial particles, where $\sigma_F/m$ likely depends on the structure of the flow as well as the relaxation time of the particle.
	In Section \ref{subsec:analysis-mle} we introduce a system for evaluating errors to better quantify comparisons to alternative noise reduction schemes, and we look at both time and frequency interpretations of the filters in Section \ref{sec:results}. 
	
	\subsection{Error analysis}
	\label{subsec:analysis-mle}
	Data filtering problems involve a fundamental trade-off. 
	A perfect filter should fully eliminate measurement noise without degrading unknown signals of interest.
	These two goals are in competition with one another - aggressive noise mitigation often leads to collateral damage in the form of signal degradation.
	Using this idea, we define two types of errors that arise in using linear filters, noting that both the MLE filter of Proposition \ref{prop:mle-solution} and local Gaussian smoothing are linear operations.
	
	Suppose that $\Phi$ represents a linear filter, and define the post filtering error as
	\begin{equation}
		\label{eqn:error-def}
		\epsilon = \Phi y - x.
	\end{equation}
	Substitution using equation \ref{eqn:meas-sum}, utilizing the linearity of $\Phi$, and rearranging terms allows us to write
	\begin{equation}
		\label{eqn:error-sum}
		\epsilon = (\Phi x - x) + (\Phi z),
	\end{equation}
	where parentheses have been added to emphasize how we intend to decompose the error.
	We define the \emph{degradation error} as
	\begin{equation}
		\label{eqn:degradation-error}
		\epsilon_x = \Phi x - x = (\Phi - I)x,
	\end{equation}
	and the \emph{pass-through noise} as
	\begin{equation}
		\label{eqn:pass-thru-noise}
		\epsilon_z = \Phi z.
	\end{equation}
	When a filter perfectly captures the true trajectory of the particle, $\Phi x = x$, the degradation error vanishes.
	Similarly, when a filter fully eliminates the noise samples, $\Phi z = 0$, the pass-through noise vanishes. Equations \ref{eqn:degradation-error} and \ref{eqn:pass-thru-noise} enable us to quantitatively compare the effectiveness of the newly derived MLE filter with other common filtering approaches. Specifically, Section \ref{sec:results} compares the new MLE filter to local smoothing using a Gaussian kernel proposed in \cite{Mordant2004} and penalized B-splines proposed in \cite{Gesemann2018}.
	
	
	\section{Methods}
	\label{sec:methods}
	This section contains details on the specific simulated trajectories used in the experimental investigation, justifications for the selection of certain underlying filter parameters, and definitions of error metrics to evaluate performance on the test set.
	For clarity, we will refer to the filtering method proposed in \cite{Mordant2004} as `Gaussian filtered', the filtering method proposed in \cite{Gesemann2018} as `B-splines filtered', and the filtering method proposed in this work as `MLE filtered'.
	\subsection{Description of the JHTDB and trajectory construction}
	\label{subsec:desc-of-experiments}
	We examined the motions of $2197$ simulated fluid particle trajectories over $2513$ time steps driven by forced homogeneous isotropic turbulence with $\mathrm{Re}_\lambda \sim 433$. 
	The DNS data was obtained from the Johns Hopkins turbulent database \cite{Perlman2007,Li2008}.
	Initially, the sampled particles were seeded uniformly within the flow.
	We chose a time step between samples of $\Delta t \approx \tau_\eta/10$. 
	At each of these time steps, we determined the position and velocity of the particles using functions included in the database \cite{Perlman2007,Li2008}.
	For both the position and velocity data, we used a 6th-order Lagrangian interpolation scheme in space. 
	For position, the integration time step was set to $0.0004$. 
	For velocity, we used a piece-wise cubic Hermit interpolation scheme in time. 
	We implemented simple finite differences of the velocity to estimate the true acceleration of each particle.
	
	\subsection{Track filtering}
	We assumed each of the $2197$ trajectories was noiseless and added simulated, normally distributed, zero-mean noise to each trajectory before applying the various filtering schemes.
	For each of the filter schemes, we optimized the filters using methods from the literature, which we expand on in this section.
	
	\begin{figure}
		\centerline{\includegraphics{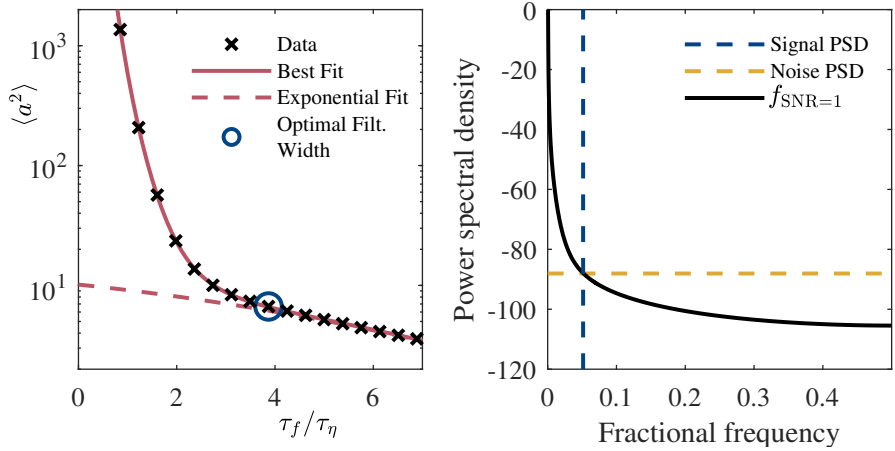}}
		\caption{Analyses used to optimize filter parameters for the various filtering methods. {\it Left}: Acceleration variance as a function of Gaussian kernel width. The solid red line is the fit of the acceleration variance from \cite{Voth2002}. The dashed red line corresponds to the exponential component of the fit. The optimal kernel width is denoted by the blue circle. {\it Right:} Power spectral density (PSD) of the particle positions is indicated by the solid black line. The yellow dashed line indicates the PSD of the added Gaussian noise and the blue dashed line indicates the fractional frequency associated with a signal-to-noise ratio of 1. Through an exhaustive search, we find that the optimal filter parameter for the B-splines method, $\lambda$, corresponds approximately to this frequency.}
		\label{fig:opt-param-curvesGauss}
	\end{figure}
	
	For the Gaussian filter method, we determined the optimal filter width, $w$, by examining the impact of the filter width on the variance of the particle acceleration magnitude, which can be seen in Figure \ref{fig:opt-param-curvesGauss} ({\it Left}).
	To compute acceleration estimates, the Gaussian kernel is twice differentiated \cite{Mordant2004}.
	The filter width $w$ has units of samples and is converted into units of time via $\tau_f = w\Delta t$. 
	For small filter widths, the variance exhibits a power law decay due to the attenuation of noise with widening averaging windows; as the width increases, the variance begins to decay exponentially because the noise mitigation is subject to diminishing returns as the window grows.
	In addition, larger filter widths increase degradation of the underlying true signal in general. 
	Therefore, the smallest filter width for which the variance begins to decay exponentially is taken to be the optimal width, shown as the blue circle in Figure \ref{fig:opt-param-curvesGauss}.
	For the given data set, this is found to be $\tau_f = 0.16$ or, equivalently, $w = 41$ samples.
	
	For the B-splines and MLE filtering methods, we determined the filter parameters $\lambda$ and $\mu$ through an exhaustive search. 
	Using this approach, we find the optimal filter parameters to be $\lambda = 236$ and $\mu = 0.0038$. 
	To validate these findings, we perform spectral analysis on the true position signal and find the frequency at which the signal strength matches that of the noise, $f_{SNR = 1}$, as shown in Figure \ref{fig:opt-param-curvesGauss} ({\it Right}). 
	This approach parallels the tuning methods performed by the TrackFit function designed by \cite{Gesemann2018}. 
	We find that the values found for both $\lambda$ and $\mu$ correspond to a signal-to-noise ratio of $SNR \approx 1$.
	For both the MLE and B-splines filters, the filters are first applied to the position data. 
	The results are then converted to accelerations using a simple central difference scheme as in equation \ref{eqn:diff-accel}.
	
	\begin{table}
		\begin{center}
			\def~{\hphantom{0}}
			\begin{tabular}{rccl}
				\textbf{Parameter}   		&   \textbf{Symbol}     & \textbf{Value} & \textbf{Units}\\[3pt]
				Number Trajectories         &   $N$                 & 2197           & n/a\\
				Sampling Rate               &   $f_s$               & 250            & Hz\\
				Noise Std Dev (position)    &   $\sigma$		    & 6.3 		     & mm\\
				Trajectory Length           &   $T$                 & 2515           & time samples\\
				Gaussian Window Length 	    &   $w$ 		        & 41 		     & time samples\\
				MLE Filter Parameter        &   $\mu$               & 0.0038         & $\text{s}^2$\\
				B-Splines Filter Parameter   &   $\lambda$           & 236            & dimensionless\\
				Taylor Reynolds Number      &   $\text{Re}_\lambda$ & 433            & dimensionless\\
			\end{tabular}
			\caption{Summary of parameters used in the experimental investigation.}
			\label{tab:exp-params}
		\end{center}
	\end{table}
	
	Once we determined the optimal filtering parameters, each filter was applied to each component of each trajectory independently; in other words the first component of position was smoothed independent of the second and the third, and vice versa, for each test trajectory.
	Table \ref{tab:exp-params} contains the prescribed quantitative parameters used throughout the experimental investigation and are consistent with the graphs shown throughout this paper. 
	
	\subsection{Performance Metrics}
	\label{subsec:error-metrics}
	We aggregated error statistics using the noiseless simulated trajectories as the definition of truth to demonstrate the effectiveness of the two techniques. For each trajectory, the true signal $x$ and measurement noise $z$ are known at each time sample, which allows us to compute the normalized total error, degradation error, and pass-through noise explicitly for all times by using equations \ref{eqn:error-sum}, \ref{eqn:degradation-error}, and \ref{eqn:pass-thru-noise}, respectively.
	We define a normalized total error at time $t$ as
	\begin{equation}
		\label{eqn:norm-tot-error}
		\overline{\epsilon}(t) = \frac{\hat{x}(t) - x(t)}{||x(t)||},
	\end{equation}
	where $\hat{x}$ is the smoothed trajectory and $||x(t)||$ denotes the vector norm of $x(t)$.
	Similarly, we normalize the degradation error and pass-through noise as
	\begin{equation}
		\label{eqn:norm-degradation-error}
		\overline{\epsilon_x}(t) = \frac{\epsilon_x(t)}{\sigma},
	\end{equation}
	and
	\begin{equation}
		\label{eqn:norm-pass-thru-noise}
		\overline{\epsilon_z}(t) = \frac{\epsilon_z(t)}{\sigma},
	\end{equation}
	respectively.
	
	The normalized total error in equation \ref{eqn:norm-tot-error} represents the difference of the smoothed trajectory to the true trajectory, relative to the true trajectory at each time.
	The normalized degradation error in equation \ref{eqn:norm-degradation-error} represents the degradation of the true trajectory under the effects of the filter, relative to the raw measurement noise.
	Finally, the normalized pass-through noise of equation \ref{eqn:norm-pass-thru-noise} represents the fraction of the measurement noise that passes through the filter and remains present in $\hat{x}$ at time $t$, relative to the raw measurement noise present at that time.
	
	Normalizing the degradation error and pass-through noise relative to the raw noise power allows us to interpret their interaction rather intuitively.
	Recall the motivation for developing these filters is to optimally reduce noise on captured data while preserving the underlying truth signal.
	A filter that diminishes noise by half but simultaneously induces signal degradation of equal magnitude is therefore not desirable.
	With this in mind, we seek filtering schemes which reduce the normalized pass-through error by more than the normalized degradation error, and we think of the normalized degradation error as the cost we incur for the pass-through noise reduction we realize. 
	
	For each trajectory, we compute three error statistics corresponding to the root mean square (RMS) error of each of the normalized errors over the course of the trajectory.
	For clarity, these are computed as
	\begin{equation}
		\label{eqn:tot-traj-error}
		E_T = \sqrt{\frac{\sum_{t=1}^T ||\overline{\epsilon}(t)||^2}{T}},
	\end{equation}
	\begin{equation}
		\label{eqn:degrade-traj-error}
		E_x = \sqrt{\frac{\sum_{t=1}^T ||\overline{\epsilon_x}(t)||^2}{T}},
	\end{equation}
	and
	\begin{equation}
		\label{eqn:pass-thru-traj-error}
		E_z = \sqrt{\frac{\sum_{t=1}^T ||\overline{\epsilon_z}(t)||^2}{T}},
	\end{equation}
	where these denote the trajectory total RMS error, trajectory degradation RMS error, and trajectory pass-through noise RMS error respectively.
	These definitions allow us to cleanly aggregate filter performance across the trajectories.
	
	Note that $E_x$ is the signal degradation cost we pay to realize the noise reduction captured by $E_z$.
	Since $E_z$ is a measure of the \emph{remaining} fractional noise power then $1 - E_z$ is a measure of the noise that was mitigated by filtering.
	Thus the ratio
	\begin{equation}
		\label{eqn:filt-payoff-rat}
		r = \frac{1 - E_z}{E_x}
	\end{equation}
	describes the noise reduction per signal degradation cost.
	Filters with $r < 1$ are inefficient as they induce more signal degradation than they reduce noise, causing them to degrade performance.
	Filters with $r > 1$ are efficient in that they improve the quality of the signal.
	Therefore we define the filter efficiency as
	\begin{equation}
		\label{eqn:filt-efficiency}
		\rho = r - 1 = \frac{1 - E_z - E_x}{E_x},
	\end{equation}
	noting that negative and positive $\rho$ values indicate inefficient and efficient filters, respectively. 
	We present the distribution of each error, as well as the filter efficiency,  based on the simulated trajectories in the following section.
	

	
	\section{Experimental Results}
	\label{sec:results}
	
	\begin{figure}
		\centerline{\includegraphics{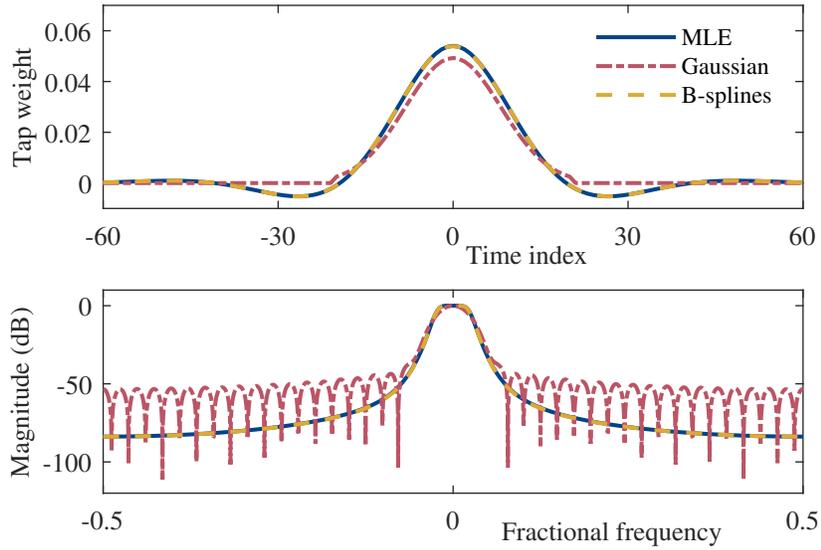}}
		\caption{Comparison of MLE, Gaussian, and B-splines filters used for position filtering with MLE filter parameter $\mu = 0.00378$, Gaussian window size of $w = 41$, and B-splines filter parameter $\lambda = 236$. Magnitudes are specified in decibels in the frequency domain relative to the greatest weight.}
		\label{fig:filter-comp}
	\end{figure}
	
	\subsection{Filter Visualization}
	We show the three optimized filters for estimating true position from noisy position data in both the time and frequency domain in Figure \ref{fig:filter-comp}. 
	Fractional frequency is defined as the ratio of the frequency to the sampling frequency of the data.
	Examination of the frequency characteristics in particular enables us to highlight key differences.
	The MLE, Gaussian, and B-splines techniques each behave qualitatively as low pass filters with individual characteristics.
	Notably however, the Gaussian filter exhibits ringing behavior across higher frequencies in comparison to the smoother decay of both the MLE and B-splines filters.
	The lobes of the Gaussian filter are accentuated by the decibel scaling of the vertical axis in Figure \ref{fig:filter-comp},
	and their presence allows high frequency noise to pass into the filtered signal.
	
	\begin{figure}
		\centerline{\includegraphics{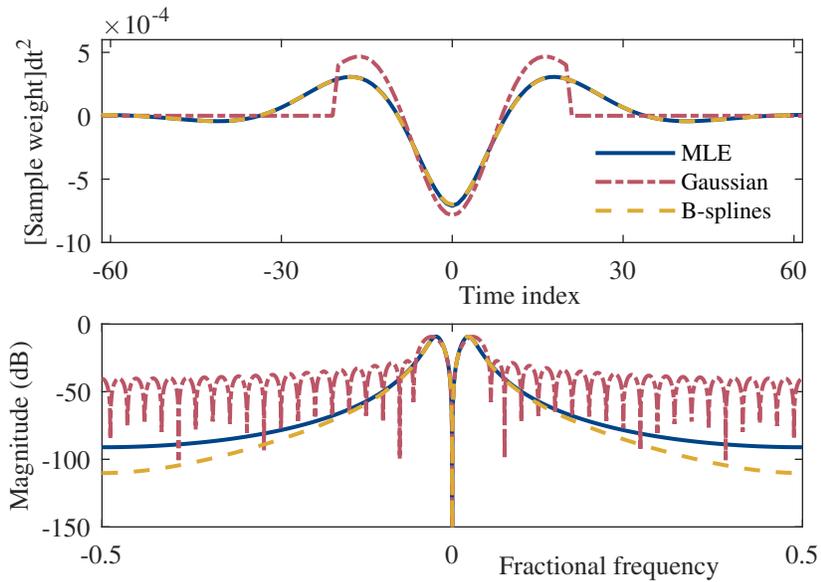}}
		\caption{Comparison of the MLE, Gaussian, and B-splines filters used to estimate acceleration. Filter parameters for each method match those used in Figure \ref{fig:filter-comp} and listed in Table \ref{tab:exp-params}. Magnitudes are specified in decibels in the frequency domain relative to the greatest weight.}
		\label{fig:acceleration-kernels}
	\end{figure}
	
	The time and frequency representations of the acceleration estimation techniques for each filtering method are depicted in Figure \ref{fig:acceleration-kernels}. 
	The sharp falloff of the Gaussian filter in the time domain is due to the discrete choice of local window length indicating the truncation point.
	Similar to the position smoothing filter it is derived from, the Gaussian filter exhibits ringing behavior allowing the passage of high frequency noise into the acceleration measurements in contrast to the smooth decay of the MLE and B-splines filters.
	In Section \ref{sec:results}, we develop error definitions that are then used to quantify this observation and demonstrate that the Gaussian filter both causes more damage to signals of interest and also fails to mitigate noise as effectively as the MLE and B-splines filters.
	
	
	\begin{figure}
		\centerline{\includegraphics[scale=1.00]{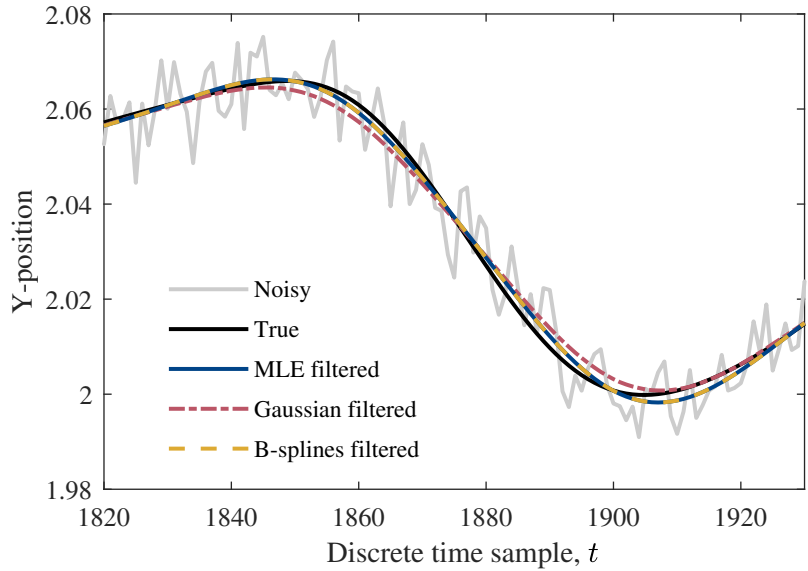}}
		\caption{Example of a smoothed position trajectory using the MLE filter, the Gaussian kernel, and B-splines filter. 
			All filters operate on the noisy data (grey) and aim to match the true position (black).}
		\label{fig:samp-traj}
	\end{figure}
	
	\subsection{Performance and Discussion}
	\label{subsec:perform-ex}
	
	We first examine the performance of each of the filtering techniques in regards to estimating the true position from noisy position data.
	Figure \ref{fig:samp-traj} shows the true $y$-position of a sample trajectory over a short time window along with the noisy data and filtered estimates.
	The true $y$-position reaches a local maximum near index $1855$ that is difficult to discern in the noisy signal.
	While all filtering methods reveal the presence of a local maximum, the Gaussian method fails to accurately capture the greatest value the true signal achieves.
	Conversely, the MLE and B-splines methods very nearly reach the true value of the local maximum while remaining closer to the truth signal over the entire sample window, including where the filtering methods more closely capture the presence of a local minimum near index $1900$.
	A Gaussian smoothing process is unable to reproduce local minima or maxima, and this is one mechanism by which the signal degradation error of the Gaussian method is greater than that of the MLE and B-splines methods.
	
	\begin{table}
		\begin{center}
			\def~{\hphantom{0}}
			\begin{tabular}{lccc}
				\textbf{Metric}   		       &   \textbf{MLE Filter}  &\textbf{B-Splines Filter} & \textbf{GS Filter}\\
				RMS Degradation Error ($E_x$)  &   0.1134               & 0.1138                  & 0.2108\\
				RMS Pass-Through Error ($E_z$) &   0.3693               & 0.3690                  & 0.4009\\
				RMS Total Error ($E_T$)        &   0.0702		        & 0.0701                  & 0.0997\\
				Filter Efficiency ($\rho$)     &   7.9746               & 7.9457                  & 2.0169\\
			\end{tabular}
			\caption{Comparison of average filter performance metrics over trajectory test set.}
			\label{tab:exp-means}
		\end{center}
	\end{table}
	
\begin{figure}
		\centerline{\includegraphics{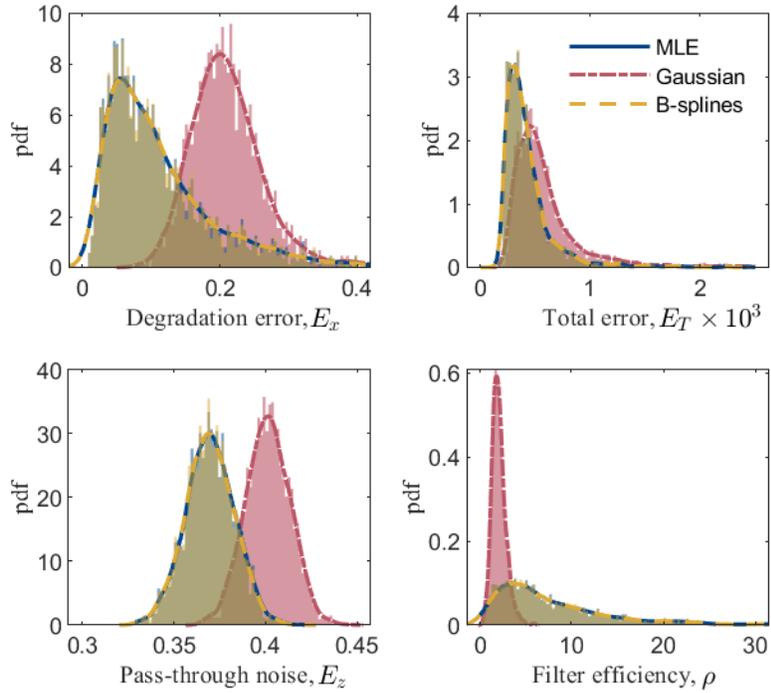}}
		\caption{Probability density function (pdf) comparisons of degradation error ({\it top left}, equation \ref{eqn:degrade-traj-error}), the total error ({\it top right}, equation \ref{eqn:tot-traj-error}), pass-through noise ({\it bottom left}, equation \ref{eqn:pass-thru-traj-error}), and filter efficiencies ({\it bottom right} equation \ref{eqn:filt-efficiency}) of the MLE, Gaussian local averaging, and B-splines filters over all experimental trajectories.}
		\label{fig:2error-hist}
		\label{fig:total-error-efficiency-hist}
	\end{figure}

	We show that both the MLE and the B-splines schemes outperform the Gaussian local smoothing by all performance metrics developed in Section \ref{subsec:error-metrics} for our experimental set.
	Table \ref{tab:exp-means} contains the average values of the performance metrics over all $2197$ trajectories and $2515$ time steps for each of the filters.
	We find that the MLE filter causes less than $50\%$ of the RMS degradation error when compared to the Gaussian smoothing method.
	Moreover, in conjunction with the reduced pass-through noise observed in the application of the MLE and B-splines filters, this leads to substantially improved filter efficiencies.
	The average efficiency of these filters is nearly four times greater than that of the Gaussian smoothing.
	The distributions of the degradation error and pass-through noise over the set of simulated trajectories are shown in Figure \ref{fig:2error-hist}.
	Additionally, the total error observed using the MLE or B-splines schemes is nearly $30\%$ less than that observed using the Gaussian smoothing scheme.
	The distributions of total error and filter efficiency are shown in Figure \ref{fig:total-error-efficiency-hist}.
	
	The MLE, Gaussian , and B-splines smoothing methods produce $E_x$ and $E_z$ distributions that are monomodal.
	For both errors, the MLE and B-splines methods achieve a substantially smaller mean compared to the Gaussian smoothing, with values shown in Table \ref{tab:exp-means}. 
	The width (second moment) of the error distributions are relatively insensitive to the present choice of smoothing method.
	
	Similar to the decomposed error, the distribution of the total error $E_T$ and filter efficiency $\rho$ are also monomodal for all filtering methods.
	MLE and B-splines outperform the Gaussian smoothing method in terms of the total error and efficiency, again as indicated by the mean values shown in Table \ref{tab:exp-means}.
	The MLE and B-splines methods exhibit a narrower distribution in the total error compared to the Gaussian smoothing method, demonstrating that the MLE and B-splines filters are more consistent, rarely achieving large total error values.
	Unintuitively, the distributions of the filter efficiency for the MLE and B-splines methods are broader than for Gaussian smoothing - this indicates the MLE and B-splines filters somewhat frequently achieve very high efficiency values.
	\begin{figure}          \centerline{\includegraphics{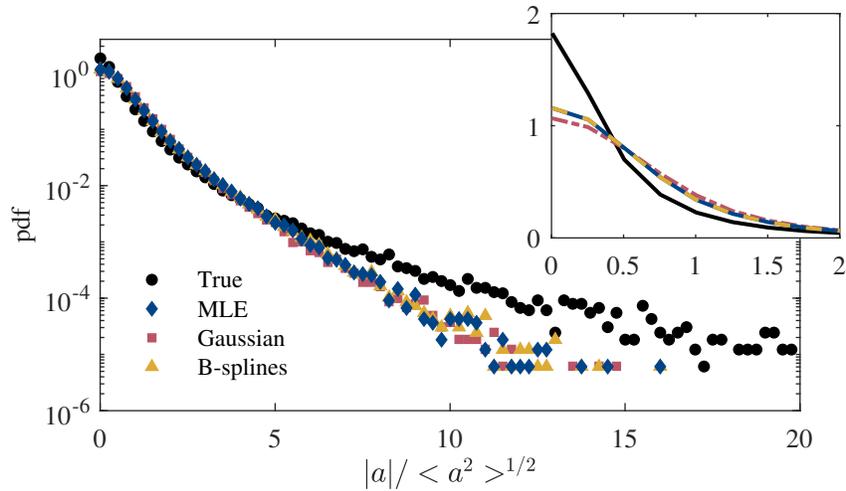}}
		\caption{Probability density functions (pdfs) of absolute accelerations obtained for each of the filtering methods. The pdfs are compared to the true accelerations (black), which are found via central differencing on the noiseless velocity.}
		\label{fig:abs-acceleration-pdfs}
	\end{figure}
	
	Finally, we compare the filters with respect to their impact on the acceleration pdf. 
	Figure \ref{fig:abs-acceleration-pdfs} shows the normalized acceleration pdf for all three filtering methods compared to the noiseless, true acceleration pdf. All three methods tend to suppress the tails of the acceleration pdf to an equal extent.
	The inset of Figure \ref{fig:abs-acceleration-pdfs} highlights the differences in estimating the core of the pdf. 
	While all methods tend to underestimate the core of the pdf, both the MLE and B-splines methods perform better than the Gaussian method.
	

	\section{Conclusions and Future Work}
	\label{sec:conclusion}
	We have developed a new, physically-grounded data smoothing scheme using Maximum Likelihood Estimation techniques and examined this new technique's ability to estimate both the position and acceleration of tracer particles in a simulated, turbulent flow. 
	We found a significant reduction in the error generated by this technique when compared to the commonly used Gaussian kernel method, indicating increased noise mitigation and decreased signal degradation. 
	Additionally we compare our new scheme to the B-splines method and find, for the current assumptions implemented in the MLE scheme, the two filters perform similarly.
	
	While the performance of the MLE and B-splines filters are nearly identical under the models utilized in this work, the probability models implemented within the MLE filter can be readily extended to much greater classes of systems exhibiting different stochastic processes and measurement noise.
	The consistency in experiments between the two techniques stems from the model of stochastic forcing we have utilized,
	it being equivalent to penalizing the third temporal derivative.
	The spline filters necessarily penalize the third derivative in the way they are constructed using Gesemann's technique, inducing the near identical behavior between the two schemes.
	The Gesemann scheme utilizes this construction independent of underlying flow statistics, whereas the MLE filter is constructed adaptively based on the stochastic physics.
	This suggests that if the probability distributions governing the fluid particle motion were better modeled (for example, an improved estimate of the variable $u(t)$ in equation \ref{eqn:stoch-force}), then the MLE filter performance would adapt accordingly.
	When different distributions are better suited to describe the flow physics, then improved performance should be realized by the adapted MLE filter in comparison to the B-splines filter.
	Exploration of optimal stochastic process models for flows of interest are an ongoing effort of future work, and we anticipate that future developments should highlight the added benefits of MLE filtering systems when compared to the B-spline filters.
	
	\section{Appendix}
	\label{sec:appendix}
	\subsection{Proof of Proposition \ref{prop:mle-solution}}
	\label{subsec:appen-proof}
	
	\begin{proof}
		\begin{equation*}
			\underset{x}{\min} \hspace{0.25cm} \frac{1}{2 \sigma^2}||y - x||^2 +\frac{m^2}{2 \sigma_F^2}||\Delta A x||^2
		\end{equation*}
		is equivalent to
		\begin{equation*}
			\underset{x}{\min} \hspace{0.25cm} \frac{1}{2}(x-y)^T(x-y) +\frac{\mu^2}{2} x^T (\Delta A)^T \Delta A x
		\end{equation*}
		with $\mu = \frac{m \sigma}{\sigma_F}$.
		Computing the gradient of the objective with respect to $x$, and setting it equal to zero, yields the optimization condition
		\begin{equation*}
			(x-y) +\mu^2 (\Delta A)^T \Delta A x = 0.
		\end{equation*}
		This is simplified through standard manipulations to the expression
		\begin{equation*}
			(I + \mu^2 A^T \Delta^T \Delta A)x = y.
		\end{equation*}
		The matrix $\mu^2 A^T \Delta^T \Delta A$ is positive semi-definite, and therefore the matrix sum $I + \mu^2 A^T \Delta^T \Delta A$ is positive definite and thus permits an inverse in general.
		Therefore the minimizing $x$ of the objective is generally written in terms of this inverse as
		\begin{equation*}
			x = (I + \mu^2 A^T \Delta^T \Delta A)^{-1}y.
		\end{equation*}
	\end{proof}
	
	\bibliographystyle{plainnat.bst}
	\bibliography{bibliography.bib}	
\end{document}